\documentclass[draftclsnofoot,onecolumn,11pt]{IEEEtran}
\ifCLASSINFOpdf
   \usepackage[pdftex]{graphicx}
\else
\fi
\usepackage[cmex10]{amsmath}
\usepackage[tight,footnotesize]{subfigure}
\usepackage{amsfonts}
\usepackage{graphics} % for pdf, bitmapped graphics files
\usepackage{setspace}
\usepackage{amssymb}
\usepackage{epsfig} % for postscript graphics files
\usepackage{stfloats}
\begin{document}
\title{Aggressive Congestion Control Mechanism for Space Systems}

\author{Jingjing~Wang, Chunxiao~Jiang,~\IEEEmembership{Member,~IEEE}, Haijun~Zhang,~\IEEEmembership{Member,~IEEE}, Yong~Ren,~\IEEEmembership{Member,~IEEE}, and~Victor~C.~M.~Leung,~\IEEEmembership{Fellow,~IEEE}% <-this % stops a space
\thanks{J. Wang, C. Jiang, and Y. Ren are with the Department
of Electronic Engineering, Tsinghua University, Beijing, 100084, China. E-mail: wang-jj14@mails.tsinghua.edu.cn, chx.jiang@gmail.com, reny@tsinghua.edu.cn.}
\thanks{H. Zhang and V. C. M. Leung are with the Department of Electrical and Computer Engineering,
The University of British Columbia, Vancouver, BC V6T 1Z4, Canada
(e-mail: dr.haijun.zhang@ieee.org; vleung@ece.ubc.ca).}
}

\maketitle

\begin{abstract}
%\boldmath
How to implement an impeccable space system-of-systems (SoS) internetworking architecture has been a significant issue in system engineering for years. Reliable data transmission is considered one of the most important technologies of space SoS internetworking systems. Due to the high bit error rate (BER), long time delay and asymmetrical channel in the space communication environment, the congestion control mechanism of classic transport control protocols (TCP) shows unsatisfying performances. With the help of existing TCP modifications, this paper contributes an aggressive congestion control mechanism. The proposed mechanism is characterized with a fast start procedure, as well as the feedback information to analyze network traffic and with a link terminating processing mechanism, which can help to reveal the real reason of packet loss, and maintain the size of congestion window at a high level. Simulation results are shown in the end to verify the proposed scheme.
\end{abstract}


% \fi
\IEEEpeerreviewmaketitle

\section{Introduction}
% The very first paper is a 2 line initial drop paper followed
Space system of systems (SoS) \cite{jamshidi2011system} is becoming a key consideration in the space communication, navigation, earth observation, etc.
The space SoS is defined as a network of assets on the Earth, in the orbit around the Earth, in the orbit around the solar system bodies, and on the surface of solar system bodies that is inter-connected and/or inter-operated to perform a mission, and/or provide services that cannot be performed by monolithic space systems alone \cite{bhasin2008communication}.
However, current space information systems are basically operated in a point-to-point pattern between the control center and the spacecrafts, which are only adequate for the individual space missions rather than collaborative missions. In other words, the ground network, space network and the deep space network evolve independently and focus on their individual communication regimes.
To address these issues, the National Aeronautics and Space Administration (NASA) has developed an integrated space system that integrates the Internet protocols, routers and interfaces into space
networks \cite{hu2013aggregation} \cite{bhasin2008architecting}. The space system, which is also called space Internet, enables large quantities of collaborative operations of those networks. Therefore, the integrated space Internet is essentially an SoS.

With more frequent and complex aeronautics and astronautics missions recently, as one of the most important technologies of space SoS internetworking systems, reliable data transmission has drawn more and more attention. It is because that costs and risks grow with the increasing of the number of links and cross-links. Therefore, in order to reduce the system risks and lower the design costs, it is necessary to establish a reliable and efficient data transmission protocol frame in space Internet. Considering the challenging communication environment in the space networks, the major differences between the space network and the ground network are summarized as follows \cite{urke2011transport}.
a) High BER: Due to the multilevel atmosphere, uncertain weather conditions and the high-speed movement of communication terminals, typical BER in the satellite communication is from $10^{-7}$ to $10^{-2}$.
b) Long time delay: Because of the long communication distance, the round trip time (RTT) is beyond imagination. In most cases, the propagation delay from the ground station to the low earth orbit (LEO) spacecraft is about 25ms, 110ms to the medium earth orbit (MEO) and 250ms to geostationary earth orbit (GEO). Even worse, in terms of the interplanetary network, the one-way propagation delay is up to seconds even minutes.
c) Asymmetric channel: In the space system, the ration of downlink bandwidth to uplink bandwidth is over $1000:1$.
d) Intermittent connection: The high-velocity motion of spacecrafts makes the communication terminals be out of sight periodically. During this period, connections are interrupted and communications are suspended.

The aforementioned challenges in space systems make it difficult to keep using the ground Internet transport control protocols to realize reliable data transmission. In \cite{fall2011tcp}, the classical TCP was a connection-oriented, duplex and reliability-ensuring protocol. It guaranteed the reliable data transmission mainly by the congestion control mechanism. TCP-Tahoe \cite{wang1992eliminating} and TCP-Reno \cite{padhye2000modeling} were the two successful versions based on scaling-windows applied in the Internet protocol structure. However, the performances of both versions were deficient and unsatisfying in the space communication environment, where it is the high BER instead of network congestion that leads to the packet loss.
Many other modified transport control protocols were also proposed for space systems. In \cite{wang2009protocols}, the authors summarized the existing modified protocols into three categories. The first category allowed the modification only at the end terminals, like TCPW \cite{casetti2002tcp}. In the second category, both transport control mechanism and intermediate nodes were improved. In this case, the TCP connection was usually divided into some sub-connections with the methods of TCP-splitting and TCP-spoofing. In the third category, the protocols could achieve the reliable data transmission function in the convergence layer
of delay-tolerant network (DTN) \cite{sun2013performance} model. In the following, this paper surveys and introduces the existing modified transport control protocols and correspondingly lists their limitations \cite{wang2009protocol}.

In Table \ref{Table.1}, we summarize the advantages and disadvantages of classical transport protocols \cite{taleb2006refwa} \cite{marchese2004petra} \cite{luglio2004board}. Among these multifarious modified protocols, some were in the testing phase emulating the real scenarios, while others were merely in the numerical analysis phases. Moreover, it is worth mentioning that DTN-based space technologies have attracted more researchers' attention in recent years. In \cite{yu2013dtn}\cite{hu2014memory}\cite{yu2015performance}, the authors proposed some data transmission protocols based on DTN, especially for deep-space communications. Moreover, the DTN-based data transmission protocols obtained a superior performances against TCP-based data transmission protocols under the long-delay and asymmetrical channel. Because TCP-based transmission protocols rely on the timely feedback of acknowledgments(ACKs), while the long-delay feedback and constrained uplink capacity result in a large number of ACKs to be lost. However, the operation of DTN-based data transmission protocols depend heavily on the bundle protocol to form a store-and-forward overlay network, which needs a cross-layer design and complex configuration in the terminals even the routers. Furthermore, the connectionless property of DTN-based protocols brings great challenges in the security issues of information transmission.
Consequently, there is not a comprehensive or efficient version towards space network transport control protocols, let alone a space-ground integrated protocol stack.
Therefore, we need urgently a simple, compatible and reliable congestion control mechanism. Furthermore, from the related work, we can summarize that the modifications should observe the following rules:
a) The redundancy of a data segment should be compressed as much as possible.
b) The congestion control mechanism and error control mechanism can adapt to the long-delay, high BER, asymmetry and intermittent interruption channel environment.
c) Take full advantage of the gateways or routers to forward data, if necessary. To the greatest extent, we should keep the original routing protocol unchangeable or less changeable.

Therefore, to address the aforementioned issues, we propose an aggressive congestion control mechanism, aiming at obtaining a higher link utilization rate and a larger throughput in data transmitting. This mechanism should be configured based on the traditional TCP protocol architecture and the data packets are transparent over intermediate nodes. Moreover, it is appropriate for the space communication environment network with comprehensive packet loss rate under 5\% as well as low end-to-end latency, while a large end-to-end delay and frequent connection interruptions are tolerant in the DTN protocols. To the best of our knowledge, the congestion control mechanism can well accommodate to the space communication environment under the hypothesis mentioned above. Besides, the proposed mechanism has the characteristics of simplicity, low cost, strong operability and easy implementation. Moreover, it obviously improves the link utilization and has a commendable performance in robustness and compatibility. The reason of calling it ''aggressive'' is that the congestion control mechanism recovers the size of sending window rapidly, maintains the congestion control window at a high level and has a remarkable performance in the high BER environment.

The rest of this paper is organized as follows. We first propose and describe an innovative congestion control mechanism for space systems in Section II. In Section III, we demonstrate the simulation results and some performance analysis. Finally, conclusions are drawn in Section IV.

\begin{table}
\centering
\caption{A summary on transport control protocols for space network}
\label{Table.1}
\begin{tabular}{||c|c|c||}
  \hline
 Protocols  &  Characteristics &  Limitations \\
  \hline
TCP-Peach& Better for long delay & Ignore high BER \\
  \hline
TCP-W& Check network traffic& Useless in long delay \\
  \hline
TP-Planet& Establish two states & Complex, incompatible \\
  \hline
XCP& IP record network condition & Split TCP connection \\
  \hline
PETRA& Intermediate node to forward & Lack of security \\
  \hline
REFWA& Better efficiency and fairness  & Ignore high BER \\
  \hline
SCPS-TP& Multiple improvement & Complex, incompatible \\
  \hline
BP& Cope with DTN & Cross-layer design  \\
  \hline
LTP& Hierarchical data transmission & Cross-layer design \\
  \hline
\end{tabular}
\end{table}

\section{Congestion Control Mechanism for Space Systems}
 In view of the existing modification methods which either lacked the compatibility with the ground Internet or excessively reduced the sending window, we propose a new ''aggressive'' modification mechanism with the following motivations. First, we should take full advantage of the characteristic of large bandwidth-delay produce (BDP) in the space communication environment. A feasible method tries to increase the sending window and to change the classical TCP slow start phase. Second, cutting the congestion threshold down cursorily squanders the channel resources. We should make use of the historical information to estimate the network traffic condition and smoothly reduce the congestion threshold. Third, the proposed modification can well manage the abrupt link interruption due to the high speed movement of aircrafts.

 \emph{Fast start}: Given the receivers feeding back one acknowledgement (ACK) for each successfully received data segment, according to the principles of TCP slow start procedure, it needs $t_{ss}$ to reach the stable transmission rate \emph{B}. $t_{ss}$ can be calculated by
\begin{equation}\label{1}
t_{ss}\approx RTT\cdot[1+log_2(B\cdot RTT/l)],
\end{equation}

 \noindent where $\emph{RTT}$ is the round trip time of each segment, which is about 50ms, 250ms and 550msin the scenario of LEO, MEO and GEO, respectively. $l$ equals 1KB in this paper, which represents the average bit length of data segments. If stable transmission rate $B$ is 10Mbps, according to equation (\ref{1}), in the GEO scenario, $t_{ss}$ is up to 5.73s; if the stable transmission rate $B$ can reach 100Mbps, under this assumption, the value of $t_{ss}$ is 7.91s in the same scenario. Obviously, it takes too long to meet the need of general communication missions. A simple substitution of slow start named fast start is proposed in this paper. In the fast start stage, when senders deliver a data segment, they transmit an empty segment simultaneously. The empty segment containing a compressed TCP header motivates the receivers to feed back ACKs, which largely accelerates the size of congestion window (\emph{cwnd}). The size of senders' \emph{cwnd} grows as the speed of $3^n$ rather than $2^n$. This speed-up mechanism is transparent to routers and receivers. In this respect, there is no need to modify other nodes in the network, which simplifies the modification operation and maintains a good compatibility. Therefore, fast start is compatible with the real ground Internet and space network protocol stacks.

  \emph{Adaptive congestion control mechanism}: When receiving 3 duplicate ACKs, instead of halving the \emph{cwnd}, senders utilize \emph{RTT} carried by ACKs to analyze the real reason of packet loss \cite{fu2003tcp}. For different reasons such as high BER or network congestion, they take different actions. For making accurate judgements, we should collect and analyze some parameters in the network.
 Let $\sigma$ represent the variable of throughput, which can be calculated as follows:
 \begin{equation}\label{2}
\sigma=(Expected-Actual)\cdot base\_RTT,
 \end{equation}
where $base\_RTT$ is the reference standard of \emph{RTT} and variable $Expected\!\!=\!\!cwnd/base\_RTT$. It is notable that parameter $Actual$ should be calculated accurately, because it directly reflects the present state of the network. However, the network states possess continuity and they can not change tempestuously. In consequence, the space system can be considered having memory characteristic. In order to obtain the accurate value of $Actual$, senders need to record the historical observation value $RTT_i$ and recording time $T_i$, where $i$ shows the different measure moments. Then $Actual$ can be determined by the following expressions:
  \begin{equation}\label{3}
   \begin{aligned}
&Actual=cwnd/\widetilde{RTT}\\
&\widetilde{RTT}=\sum_{i=1}^{n}A_{i}e^{-\tau}RTT_{i},\\
   \end{aligned}
  \end{equation}
where attenuation index $\tau=T_{n}-T_{i}, i=1,2,...n$, $T_n$ represents the present moment, and $T_1$ is the moment when the terminals establish the connection. $A_{i}$ is the normalization coefficient.

Furthermore, we set a threshold of $\sigma$ with variable $\beta$. As mentioned above, $\sigma$ approximately reflects the variation per $base\_RTT$ of the \emph{cwnd} against expected value. During a full $base\_RTT$, the losing of three data packets is indeed a small probability event under the condition of normal connection without network congestion. In this article, let congestion threshold $\beta$ be a constant valued 3 in this paper.
When $\sigma\leq\beta$, we regard the packet loss as a result of random bit error. Thus, the sender retransmits the lost packet immediately. Meanwhile, the congestion threshold (\emph{ssthresh}) maintains unchangeable and congestion window increases by three data segments due to the three duplicate ACKs received successfully, i.e.,
  \begin{equation}\label{4}
   \begin{aligned}
&ssthresh=cwnd \\
&cwnd=ssthresh+3.\\
   \end{aligned}
  \end{equation}
On the other hand, if $\sigma>\beta$, we regard the packet loss as a result of network congestion. At this time, the sender retransmits the lost packet, while the \emph{ssthresh} and \emph{cwnd} are calculated as the following equations:
\begin{equation}\label{5}
   \begin{aligned}
&ssthresh=cwnd\times k \\
&cwnd=ssthresh+3 \\
&k=\frac{\beta}{\sigma}\frac{base\_RTT}{\widetilde{RTT}},
   \end{aligned}
  \end{equation}

\noindent where the threshold control coefficient depends on the degree of network traffic congestion. In this way, $ssthress$ can be reset to a reasonable value.
When the retransmission timer times out, similar to the TCP-Reno mechanism, senders halve the \emph{ssthresh} and set \emph{cwnd} to 1.

 \emph{Congestion windows maintaining}: The high speed movement of satellites leads to a high probability of link interruption. An abrupt increasing of \emph{RTT} causes a leap of $\sigma$. To make matters worse, the \emph{cwnd} will shrink persistently until the date transmission link rebinds. Under this circumstance, we propose a congestion window maintaining mechanism. First, we should estimate the distance of the whole TCP connection. In \cite{pan2011improvement}, the instantaneous geocentric angle $\theta$ of two satellites is written as follows:
    \begin{equation}\label{6}
\theta=\arccos[\sin\varphi_{a}\sin\varphi_{b}+\cos\varphi_{a}\cos\varphi_{b}\cos(\psi_{a}-\psi_{b})],
   \end{equation}
where $(\varphi_{a},\psi_{a})$ and $(\varphi_{b},\psi_{b})$ are the sub-satellite point longitude-latitude coordinates of satellite $a$ and $b$. Then the distance of communication link $d$ can be calculated via:
\begin{equation}\label{7}
d\!=\!\sqrt{(r+h_{a})^{2}+(r+h_{b})^{2}-2(r+h_{a})(r+h_{b})\cos\theta},
\end{equation}
where $r$ is the earth radius, and $h_{a}$ and $h_{b}$ are the satellite orbital altitudes. Therefore, the total link distance $D$ is the sum of each point to point distance $d_{ij}$. Then the estimated $RTT_{est}=2D/c$.
The high-speed movement of two satellites leads to a corresponding change in the instantaneous geocentric angle $\theta$. According to the distribution and the altitude of stationary orbit satellites, let us assume $0\leq\theta\leq\frac{2\pi}{3}$. Based on this hypothesis, we conservatively take $10RTT_{est}$ as the interruption threshold.
When the $RTT_{i}>10RTT_{est}$, we can judge that the data transition link breaks off because of the high speed movement.
Simultaneously, maintaining the size of $cwnd$, holding the time-out clock, and recording the transmitting state, the sender continuously dispatches detective packets. Once the communication link reconnects, the data transmission recovers.

\begin{figure}[!t]
\centering
\includegraphics[width=0.6\textwidth]{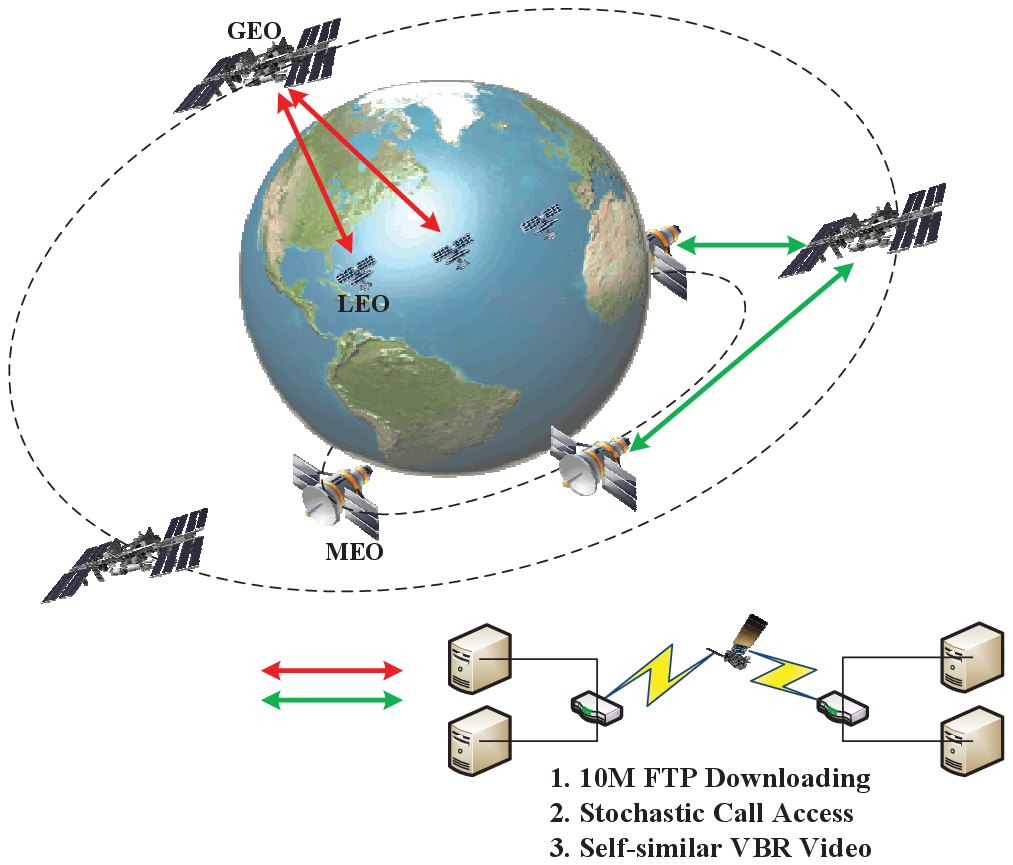}\\
\caption{\!Simulation scenario and simulation business model}
\label{Fig.1}
\end{figure}

This modified TCP congestion control mechanism can accurately distinguish the reason of packet loss, and can hold the congestion window to a large extent. In the space communication environment of a large bandwidth delay product, this aggressive modification can expand the throughput and increase the link utilization rate significantly.

\begin{figure*}
\begin{center}
\subfigure[Time-average throughput in $0.5\%$ packet loss rate]{\includegraphics[width=0.40\textwidth]{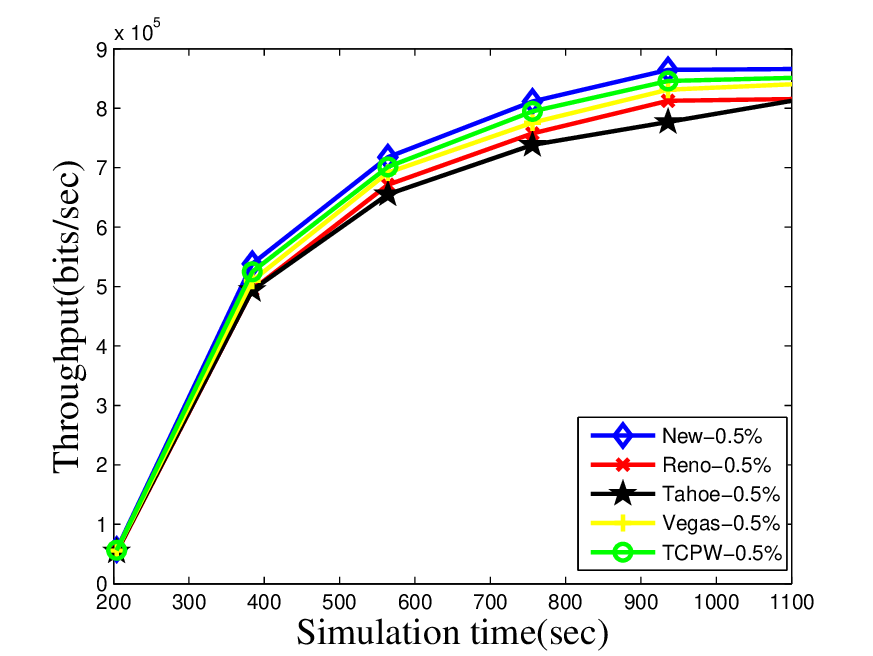}}
  \hspace{0.05in}
\subfigure[Time-average throughput in $1\%$ packet loss rate]{\includegraphics[width=0.40\textwidth]{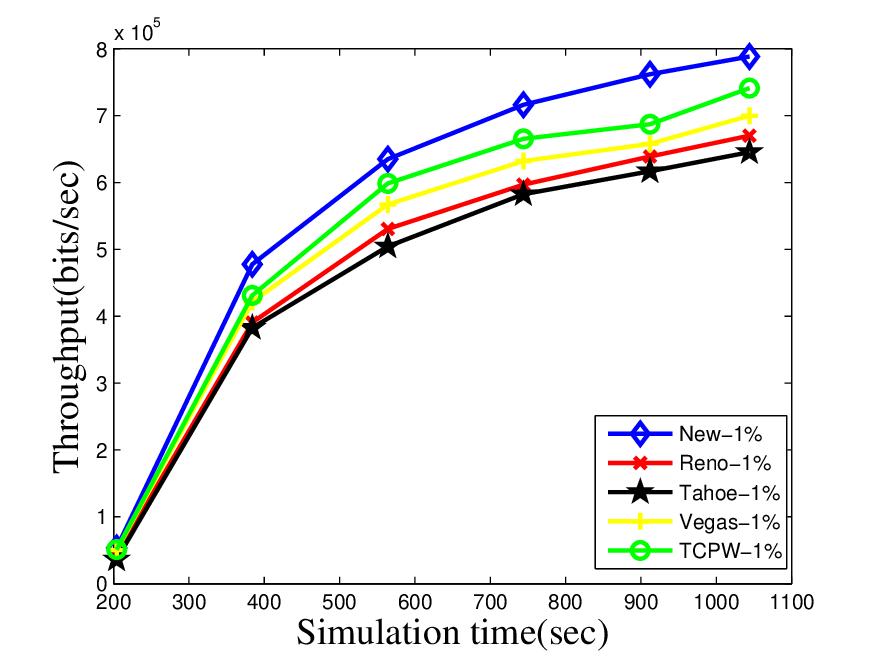}}\\
  \hspace{0.05in}
\subfigure[Real time \emph{cwnd} in $0.5\%$ packet loss rate ]{\includegraphics[width=0.40\textwidth]{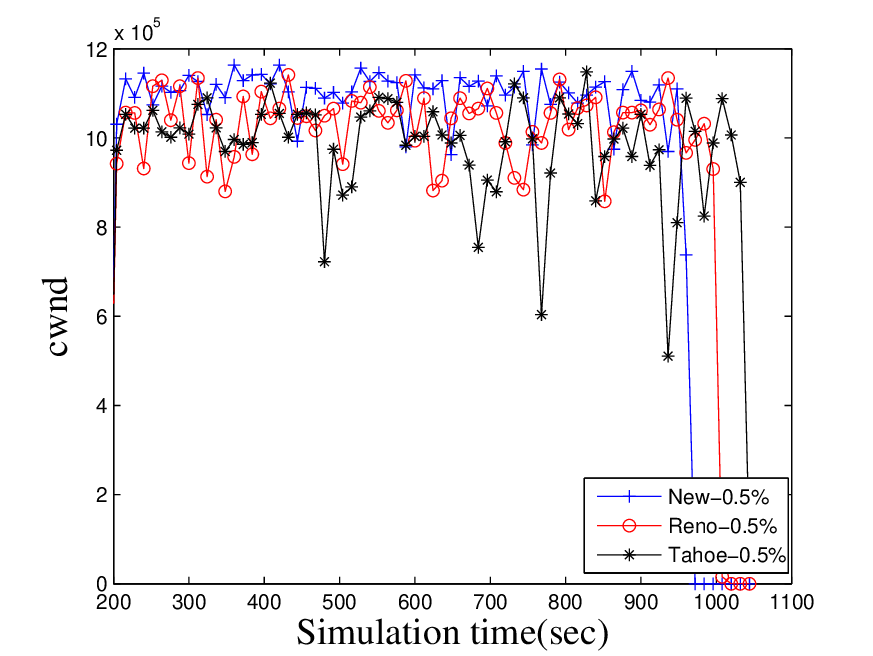}}
\hspace{0.05in}
\subfigure[Real time \emph{cwnd} in $1\%$ packet loss rate]{\includegraphics[width=0.40\textwidth]{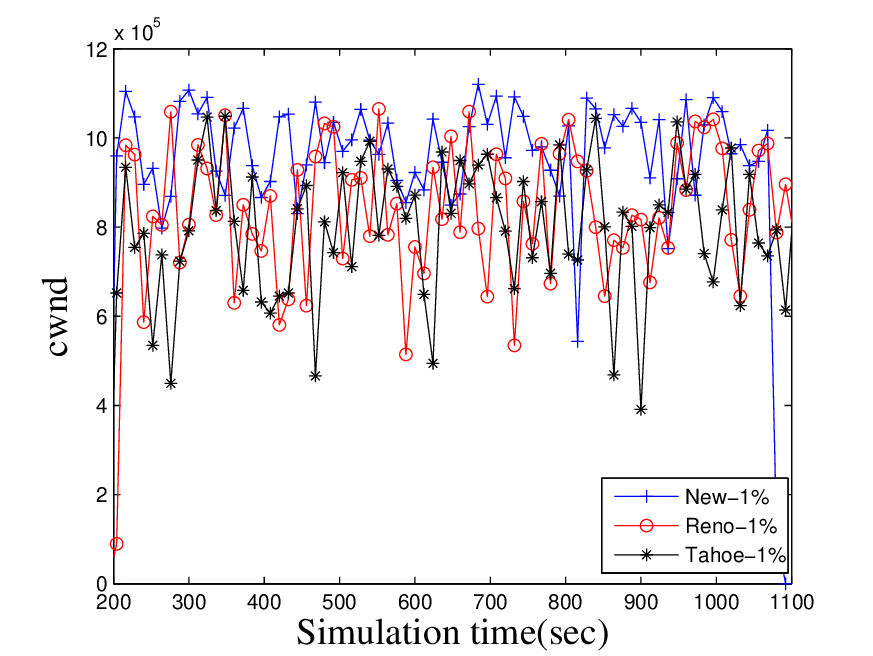}}\\
\subfigure[Time-average channel utilization in $0.5\%$ packet loss rate]{\includegraphics[width=0.40\textwidth]{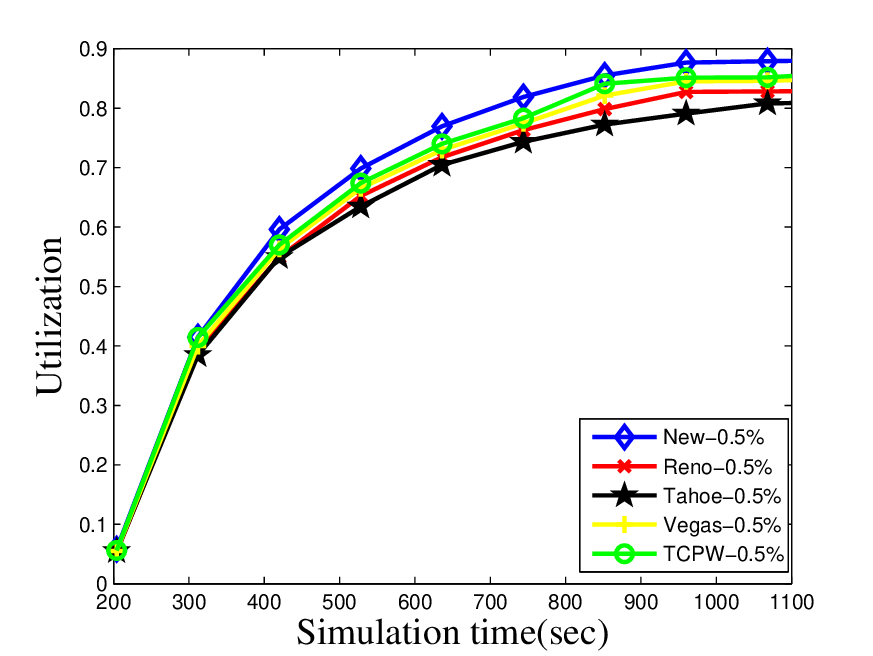}}
  \hspace{0.05in}
\subfigure[Time-average channel utilization in $1\%$ packet loss rate]{\includegraphics[width=0.40\textwidth]{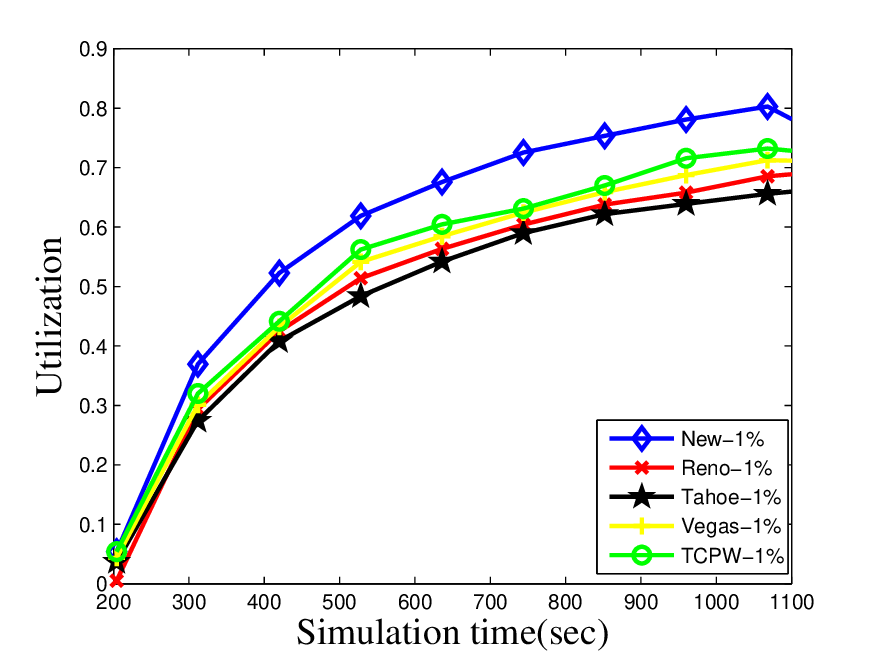}}\\
  \hspace{0.05in}
\subfigure[Time-average channel utilization in $5\%$ packet loss rate]{\includegraphics[width=0.40\textwidth]{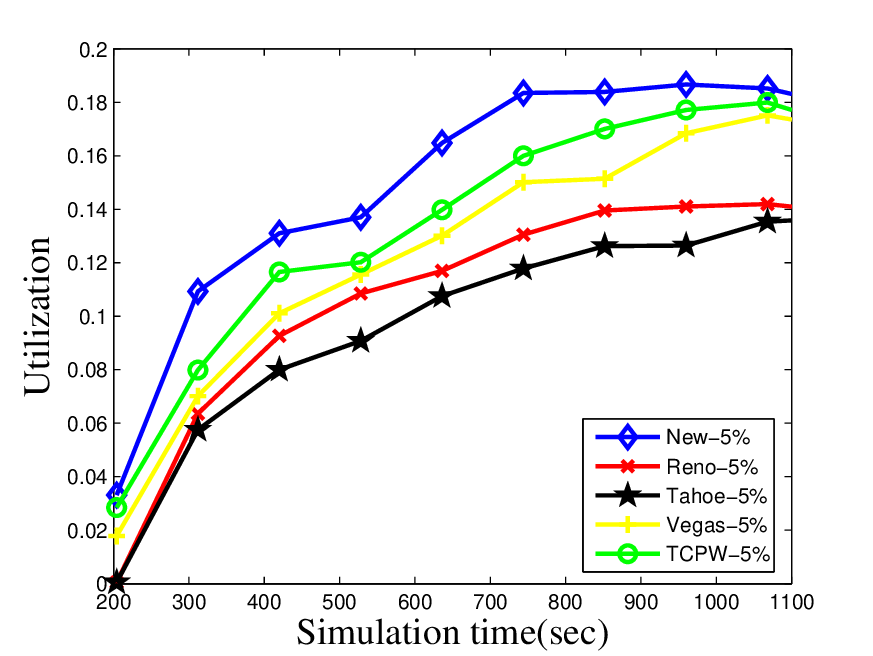}}
\hspace{0.05in}
\subfigure[Time-average throughput in VBR video scenarios]{\includegraphics[width=0.40\textwidth]{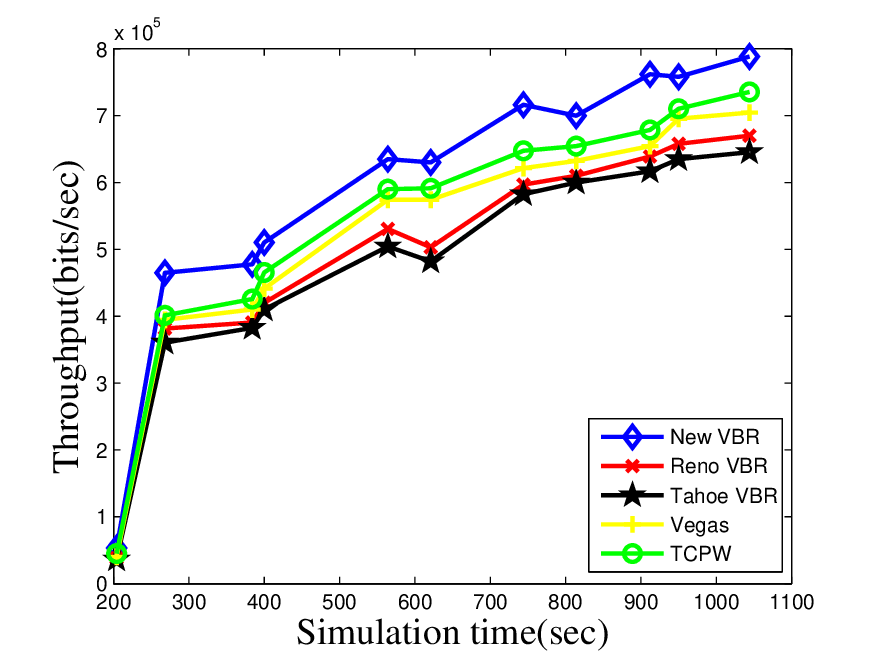}}
\caption{Simulation results in different channel situations and different business models (FTP downloading and VBR video transmitting)}
\label{Fig.2}
\end{center}
\end{figure*}

\section{Simulation Results And Analysis}

\begin{table}[!t]
\centering
\caption{Average hold-on time and new call blocking rate in the second scenario}
\label{Table.2}
\begin{tabular}{||c|c|c|c|c||}
  \hline
 Simulation Protocols  &  Average hold-on time &  New call blocking rate \\
  \hline
New TCP& 1.948ms & 0.012\% \\
  \hline
TCP-Tahoe& 2.216ms& 0.028\% \\
  \hline
TCP-Reno& 2.031ms & 0.018\% \\
 \hline
TCP-Vegas&1.998ms & 0.017\%\\
 \hline
TCPW &1.962ms  & 0.014\%\\
 \hline
\end{tabular}
\end{table}
This paper uses the network simulation software OPNET to verify the performance of the new congestion control mechanism. Fig. \ref{Fig.1} demonstrates the space network simulation scenarios. The space system is constructed by kinds of satellites in three layers denoted as GEO, MEO and LEO. The satellites in the same layer can communicate with each other by means of a relay satellite in the higher layer. The red and blue arrow lines represent the data transmission between two satellites via a relay satellite. This process of data transmission is noted in the right corner of the figure. The data through a wire link, routers, wireless link and relay satellite is received by clients from servers. All the simulations are under the circumstances with milliseconds end-to-end delay and comprehensive packet loss rate under 5\%.

In order to verify the performance of the modified protocol sufficiently, we simulate in different situations with three business models. First, we simulate three 10MB FTP persistent downloading scenarios with different packet loss rates, $0.5\%$, $1\%$ and $5\%$, respectively. The three kinds of packet loss rates reflect different round trip distance and various spatial environments in the real world. However, for obtaining an obvious result, the three packet loss rates are much higher than those in the ground Internet network. The simulation time is set to 20 minutes. During this period, clients can accomplish the FTP downloading mission from servers normally. We compare the new mechanism with classical TCP congestion control mechanisms in the expects of average throughput and link utilization. Second, the business model is purely stochastic. Fifty thousand INMARSAT telephone calls are accessed within 2 hours as a Poisson process with $0\leq\!t\!  \leq3600$ and average arriving rate $\lambda=14$. In each hold-on time, senders transmit 500 Bytes data to the receivers, and then hang off. We record and compare the average hold-on time of each call and the new call blocking rate. In the third situation, we consider the variable bit rate (VBR) video businesses which are not independent from each other. In other words, the variable bit rate (VBR) video businesses have the characteristic of long-range dependence. This self-similar property can be described by the Pareto distribution as in (\ref{9}), where $k=1$ and $\alpha=1.5$.
\begin{equation}\label{9}
  f(x)=\alpha\!k^{\alpha}/x^{\alpha+1}, x>k.
\end{equation}
During the simulation time, senders transmit large amounts of long-range dependence VBR video. And the average throughput is recorded.

From the subgraphs (a)$\thicksim$(d) of Fig. \ref{Fig.2}, we can see that the new congestion control mechanism has a better performance than others in the environment of high BER. It has a good robustness in severe circumstances. The higher the packet loss rate is, the more advantages the new mechanism has. From the average-processing throughput curves and real time $cwnd$ curves in the two situations with $0.5\%$ and $1\%$ packet loss rate, we can conclude that the receiver with the new mechanism first completes the 10MB FTP downloading mission, and maintains a higher throughput over classical Tahoe, Reno, Vegas and TCPW from the beginning to the end. The simulation results evidently reflect the superiority of the new congestion control mechanism.

%%\begin{figure}[!t]
%%\centering
%%\includegraphics[width= 3in,height=2.8in]{fig2.eps}\\
%%\caption{\!\!(a)Average throughput with $0.5\%$ and $1\%$ package loss rate.(b)Average throughput in the VBR video model.}
%%\label{Fig.2}
%%\end{figure}

%%\begin{figure}[!t]
%%\centering
%%\includegraphics[width= 3in,height=2.8in]{fig3.eps}\\
%%\caption{\!Link utilization in three scenarios with $0.5\%$, $1\%$ and $5\%$ loss rates}
%%\label{Fig.3}
%%\end{figure}

Subgraphs (e)$\thicksim$(g) of Fig. \ref{Fig.2} show the link utilization rate of each mechanism with three packet loss rates. They show that the performance of link utilization is enormously promoted with the new aggressive congestion control mechanism in high BER environment.
What is more, the new congestion control mechanism achieves the best relative performance in the scenario with $1\%$ packet loss rate against others.

From the Table \ref{Table.2}, we conclude that the new congestion control mechanism can well handle the stochastic business. With the help of the new congestion control mechanism, the hold-on time is shortened evidently. In other words, it has the least time of data transmitting and the minimum probability of new call blocking against the other TCP protocols. Moreover, the simulation results of the third scenario is shown in the subgraph (h) of Fig. \ref{Fig.2}. Obviously, the modification mechanism possesses the best performance in the average throughput against others in the VBR business model.

\section{Conclusion}

This paper presented a novel congestion control mechanism for space systems. The proposed mechanism which makes the size of \emph{cwnd} maintain a large value fully adapts to the space information environment of intermittent interruption and high BER. Because of the new congestion control mechanism being able to detect the real reason of packet loss, it can well regulate the network traffic and recover the throughput rapidly. Furthermore, the simulation results showed that the aggressive mechanism had a better performance over Reno, Tahoe, Vegas and TCPW in space communication environment. Moreover, the higher the BER is, the more obvious advantages of the new mechanism over traditional TCPs are. Compared with other well-performed modification versions, the novel modified mechanism is easier to be implemented and configured. We only need to amend the transport control protocol on the transmission side, and it is transparent to all intermediate routers and receivers.

\bibliographystyle{IEEEtran}
\bibliography{ref}


%% biography section
%%
%% If you have an EPS/PDF photo (graphicx package needed) extra braces are
%% needed around the contents of the optional argument to biography to prevent
%% the LaTeX parser from getting confused when it sees the complicated
%% \includegraphics command within an optional argument. (You could create
%% your own custom macro containing the \includegraphics command to make things
%% simpler here.)
%%\begin{biography}[{\includegraphics[width=1in,height=1.25in,clip,keepaspectratio]{mshell}}]{Michael Shell}
%% or if you just want to reserve a space for a photo:
%\end{IEEEbiography}
%% insert where needed to balance the two columns on the last page with
%%\newpage
%% You can push biographies down or up by placing
%% a \vfill before or after them. The appropriate
%% use of \vfill depends on what kind of text is
%% on the last page and whether or not the columns
%% are being equalized.
%
%%\vfill
%
%% Can be used to pull up biographies so that the bottom of Ethe last one
%% is flush with the other column.
%%\enlargethispage{-5in}
%
%
%
%% that's all folks

\end{document}